\documentclass[%
reprint,
superscriptaddress,
nofootinbib,
 amsmath,amssymb,
 aps, 
 pre,
floatfix,
]{revtex4-2}

\usepackage{graphicx}
\usepackage{dcolumn}
\usepackage{bm}

\usepackage{cancel}
\usepackage[separate-uncertainty=true,range-phrase=--,multi-part-units=single,range-units=single]{siunitx}
\usepackage{hyperref}
\hypersetup{
    colorlinks,
    linkcolor={blue},
    citecolor={blue},
    urlcolor={blue}
}
\usepackage{orcidlink}
\usepackage[capitalise]{cleveref}
\crefname{section}{Sec.}{Secs.}
\Crefname{section}{Sec.}{Secs.}
\crefname{equation}{Eq.}{Eqs.}
\Crefname{equation}{Eq.}{Eqs.}
\usepackage{etoolbox}

\DeclareMathOperator{\const}{const}

\newcommand{\kB}{k_{\text{B}}}

\newcommand{\heat}{\text{h}}
\newcommand{\matter}{\text{m}}
\newcommand{\lft}{\text{left}}
\newcommand{\rgt}{\text{right}}
\newcommand{\rf}{\text{ref}}

\newcommand{\SM}[1]{\cite[{#1}]{SM}}
\newcommand{\smDynamo}{S1}
\newcommand{\smMaster}{S2}
\newcommand{\smMirror}{S3}
\newcommand{\smSim}{S4}
\newcommand{\smAnalysis}{S5}
\newcommand{\seqThermal}{S17}

\newcommand{\seqSoret}{S19}
\newcommand{\sfigEq}{S1}
\newcommand{\sfigNN}{S2}
\newcommand{\sfigKL}{S3}
\newcommand{\stabArgon}{S1}

\makeatletter
\patchcmd{\paragraph}{-1em}{-.0001em}{}{ }
\patchcmd{\subparagraph}{-1em}{-.0001em}{}{ }
\makeatother

\begin{document}

\preprint{APS/123-QED}

\title{\textbf{Ensemble theory of thermodiffusion}}%

\author{R. Belousov\orcidlink{0000-0002-8896-8109}}
\email{roman.belousov@embl.de}
\affiliation{Cell Biology and Biophysics Unit, European Molecular Biology Laboratory, Meyerhofstraße 1, 69117 Heidelberg, Germany}
\author{J. Elliott\orcidlink{0000-0002-3703-4234}}
\affiliation{Cell Biology and Biophysics Unit, European Molecular Biology Laboratory, Meyerhofstraße 1, 69117 Heidelberg, Germany}
\author{A. Erzberger\orcidlink{0000-0002-2200-4665}}%
\email{erzberge@embl.de}
\affiliation{Cell Biology and Biophysics Unit, European Molecular Biology Laboratory, Meyerhofstraße 1, 69117 Heidelberg, Germany}
\affiliation{Department of Physics and Astronomy, Heidelberg University, 69120 Heidelberg, Germany}

\date{\today}

\begin{abstract}
 Heat flow sustained by a temperature difference generates a gradient of matter as observed, for example, in atomistic simulations of hard-core gases. We use path-ensemble theory to represent the coupled thermal and diffusive transport by dynamics on a graph. Calibrated from a single equilibrium simulation, this model accurately predicts nonequilibrium mass distribution in a hard-core gas across a wide range of applied temperature gradients. It also enables estimation of transport properties, such as the Soret coefficient, using analytical expressions.
\end{abstract}

\maketitle

Thermodiffusion couples transport of heat and matter in numerous phenomena of biological and technological relevance~\cite{Nowak2024,Magnez2022,Talbot2017I,JimnezAmaya2025,Lee2020,Geelhoed2006,Sengers2024}, including the thermophoresis, Dufour and Soret effects~\cite{Sagot2013,Piazza2008,Khler2016,Zheng2002,Platten2005,Wiegand2004}. Physical modeling of such phenomena has a long history that goes back to Maxwell and comprises theoretical approaches based on both statistical mechanics and hydrodynamics~\cite{Mayer2023,Kocherginsky2021,Burelbach2019,Falasco2016,Wrger2013,Debbasch2010,Fayolle2008,Dhont2004,Bringuier2003,Beresnev1995,Yamamoto1988,Talbot1980,Derjaguin1965,Brock1962,Waldmann1959,Epstein1929,Maxwell1879,Parola2004,Duhr2006PRL,Duhr2006PNAS,Burelbach2018,Liang2022}. In this letter we focus on the statistical origins of thermal diffusion in a single-component gas that causes redistribution of mass under a temperature gradient~\cite{Brenner2010,Teagan1968,Alofs1971,Ohwada1996,Graur2009,Wu2015,Busuioc2024}.

\begin{figure}[!t]\centering
\includegraphics[width=\columnwidth]{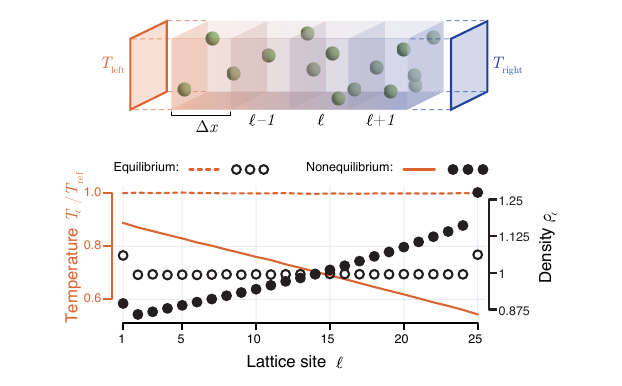}
\caption{\label{fig:intro}Molecular-dynamics simulations of a hard-core gas in a nonequilibrium steady state sustained by thermalizing walls at $T_\lft/T_{\rf} = 0.9$ and $T_\rgt/T_{\rf} = 0.5$ ($T_{\rf}$: reference temperature at equilibrium; $\kB=1$ in arbitrary units, \SM{Sec.~\smDynamo}). Histograms of particle density $\rho_\ell$ and kinetic temperature $T_\ell$ exhibit opposing gradients when $T_\lft\ne T_\rgt$ .
}
\end{figure}

In particular, we study thermodiffusion of a hard-core gas confined between two thermalizing walls at temperatures $T_\lft$ and $T_\rgt$ (\cref{fig:intro}). When a temperature difference $\Delta T = T_\rgt - T_\lft$ sustains a heat flux $\bm{j}_\heat(\Delta T)$ through the system, a thermodiffusive force generates the flow of matter. However, if matter is not exchanged with the environment, its net macroscopic current $\bm{j}_\matter$ must vanish in the nonequilibrium steady state as a consequence of the emergent density gradient $\nabla \rho$ that creates an opposing diffusive force due to Fick's law. In the linear regime $\nabla T \propto T_\rgt - T_\lft$ this condition reads~\cite{Liang2022,Parola2004,Burelbach2018,Duhr2006PRL,Duhr2006PNAS,Brenner2010}
\begin{equation}\label{eq:no_diffusion}
    \bm{j}_\matter = -D \nabla\rho - D_T \rho \nabla T \equiv 0,
\end{equation}
where $D$ and $D_T$ are the coefficients of diffusion and thermal diffusion, respectively.

The coupling between heat and matter flows in single-component fluids, which is arguably simpler than in mixtures, receives surprisingly little attention~\cite{Teagan1968,Alofs1971,Ohwada1996,Graur2009,Wu2015,Busuioc2024,Brenner2010}. Generic thermodiffusive and thermophoretic forces are commonly attributed to the differences of momentum and energy transferred between colliding molecules under a temperature gradient, which generate stresses in the direction opposing the heat flow. Kinetic theory yields estimates of the pertinent parameters that can be used to solve the Boltzmann equation or a system of hydrodynamic equations under suitable macroscopic boundary and continuity conditions~\cite{Ohwada1996,Graur2009,Wu2015,Busuioc2024}. Albeit theoretically valuable, these estimates have limited practical utility because of their approximate nature, mathematical complexity, and dependency on system properties that are hard to measure. 

The problem of thermodiffusion has also been approached by applying the principle of maximum caliber, which couples thermal and diffusive flows through the Onsager reciprocal relations~\cite{Ghosh2006,Hazoglou2015,Dixit2018}. This approach relies on a simplified convective model of energy transport, with constraints imposed on macroscopic flows across the entire system. However, as shown in Ref.~\cite[Sec.~II.D and III]{companion} and here, such constraints neglect important details of the microscopic dynamics and may fail to recapitulate actual behavior of the systems in boundary-driven nonequilibrium states.

\citet{Liang2022} proposed another simplified model of thermophoretic phenomena, which takes into account the dependency of the system's kinetic properties on its configuration. Using the framework of stochastic thermodynamics they showed that a particular interplay between the state-dependent transport coefficients may produce the Soret effect, and under different conditions even its inverse. However, how such mechanisms are related to the molecular models of matter remains unknown.

In this letter, we circumvent the above limitations by formulating a theory of thermodiffusion based on the microcanonical principle of maximum caliber (\textit{microcaliber}), which determines the probability of the system's trajectory through the calibration of dynamics at equilibrium. This theory, whose foundations are presented in the companion paper~\cite{companion}, explicitly introduces microcanonical constraints of energy and matter continuity---overt properties of molecular motion. Thereby it directly relates the system's kinetic properties to its configuration.

\begin{figure}[!t]\centering
\includegraphics[width=\columnwidth]{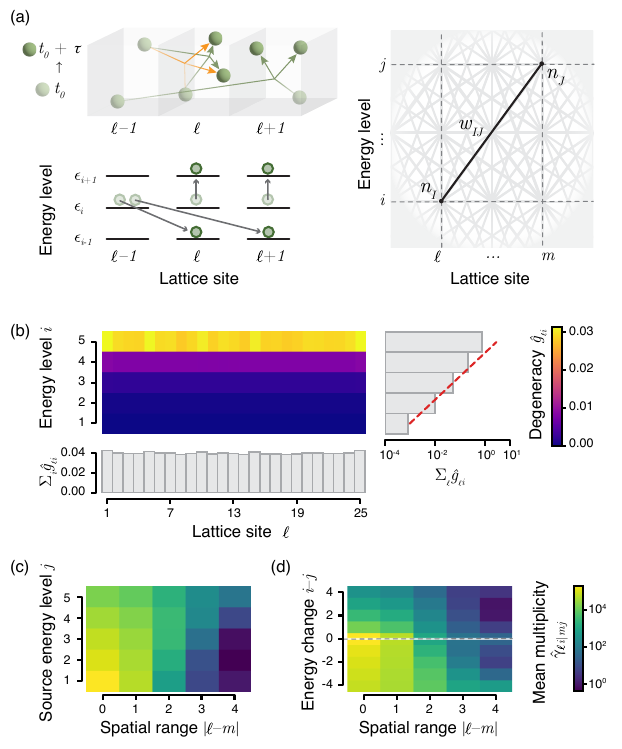}
\caption{\label{fig:graph}Graph dynamics of equilibrium MD simulations at $\beta_\rf = 1$. (a) We consider particle dynamics within discrete lattice sites $\dots, \ell -1, \ell, \ell + 1,\dots$ of size $\Delta x$ with energy levels $\epsilon_{i=\dots, i-1, i, i+1,\dots}$ (left). Orange arrows illustrate an alternative microscopic realization of a transition that contributes to its multiplicity. The particle dynamics are mapped to a graphical representation, such that each particle state $I= \{li\}$ is represented by a node, with the transitions between the source $I$ and target $J$ states described by (directed) edge weights $w_{IJ}$ (right). (b) Degeneracies $\hat{g}_{\ell i}$ projected by summation $\hat{g}_i = \sum_\ell \hat{g}_{\ell i}$ reveal a linear logarithmic trend $\ln\hat{g}_i \approx \const + 2 \ln i$ (red line), consistent with the number of single-particle energy levels growing as the area of a sphere of radius $|v| \sim i$. (c) and (d) Mean multiplicities $\hat{\gamma}_{\ell i | mj} = \hat{w}_{\ell i | mj} / \hat{g}_{\ell i}$ as the function of the transition range $|\ell - m|$ and either of the source energy level $j$ (c) or of the energy change  $i - j$ (d) show that long-range jumps become more frequent with the particle's energy, and tend to result in energy loss.}
\end{figure}

Specifically, we map the problem of thermodiffusion onto a Boltzmann gas~\cite{Sharp2015} evolving on a lattice of size $L$ with discrete energy levels and theoretically tractable microscopic graph dynamics~[\cref{fig:graph}(a)]~\cite{Burda2009,Dixit2015}. The parameters of this mapping are conveniently calibrated from a single equilibrium molecular-dynamics (MD) simulation at the reference temperature $T_\rf$~\SM{Sec.~\smDynamo}. We then analyze nonequilibrium steady states at $T_\lft > T_\rgt$, comparing our \textit{quantitative} predictions for the mass distribution and transport in a hard-core gas with simulations of the original atomistic model. We show that the thermodiffusive force emerges from the calibrated weights of the graph dynamics, without an explicit consideration of momentum transport. Remarkably, our theory accurately predicts the system properties in the MD simulations of a hard-core gas far from equilibrium, and yields a simple formula for the Soret coefficient\footnote{
    Thermodiffusion in a single-component gas, which should be distinguished from \text{Soret effect} in mixtures, allows a consistent definition of the \text{Soret coefficient}.
} $S_T = D_T / D$.



\paragraph*{Model.}---As a simplified model of a gas, we consider $N$ \textit{indistinguishable} particles, which occupy discrete energy levels $\epsilon_i$ indexed by $i=1,2,...M$, each available with degeneracies $g_{\ell i}$ at every site $\ell=1,2,...,L$ of a one-dimensional lattice with spacing $\Delta x$. Then, in the double index notation $I=(\ell i)$, the vector of occupancies $n_I = n_{\ell i}$ of all these states describes the system's configuration. Its dynamics between two time points, $t_0$ and $t_0+\tau$ is summarized by the numbers of particles $n_{IJ} = n_{\ell i | m j}\big(\tau|n_J(t_0)\big)$ moving from the energy level $J=(mj)$ to $I$.

We also introduce path multiplicities $\gamma_{IJ} = \gamma_{\ell i|mj}$ that count \textit{physically distinct} realizations of each transition $n_{IJ}$~\cite{Dixit2015}. The total number of realizations for the system's trajectory is then approximately given by the Maxwell-Boltzmann formula~\cite{companion}:
\begin{equation}\label{eq:Omega}
    \Omega_{t_0}^{t_0+\tau} \approx \prod_{IJ:\ a_{IJ}=1} \frac{\left(g_I\ \gamma_{IJ}\right)^{n_{IJ}}}{n_{IJ}!},
\end{equation}
where the product is taken over the transitions encoded in the \textit{adjacency} matrix $a_{IJ}$, whose elements equal to one when a transition $J \to I$ is allowed, and to zero otherwise.

Using the Stirling approximation, the \textit{microcanonical caliber} of the system is evaluated as~\cite{companion}
\begin{equation}\label{eq:caliber}
    \ln \Omega_t^{t_0+\tau} \approx \sum_{IJ} a_{IJ} n_{IJ} \big(\ln(g_I\gamma_{IJ}) - \ln n_{IJ} + 1\big),
\end{equation}
subject to a set of conditions for the continuity of matter
\begin{equation}\label{eq:cont_matter}
    A_J := n_J(t_0) - \sum_I a_{IJ} n_{IJ} = 0,
\end{equation}
and energy
\begin{equation}\label{eq:cont_energy}
    B_\ell := \sum_{i | m j} a_{\ell i| m j} (\epsilon_j - \epsilon_i) n_{\ell i | m j} = 0.
\end{equation}

The continuity \cref{eq:cont_matter} ensures that the number of transferred particles summed over all the target levels $I$ (including $I=J$) amounts to the number of particles $n_J(t_0)$ originally residing in the source level $J$ at time~$t_0$. This condition implies that the particles are neither created nor annihilated. It also entails
\begin{equation}\label{eq:update}
    n_I(t_0 + \tau) = \sum_J n_{IJ}\big(\tau | n_J(t_0)\big).
\end{equation}

Equation~\eqref{eq:cont_energy} balances the total energy carried over by the particles from the source level $\epsilon_j$ at the site $m$ to the site $\ell$, which at time $t_0+\tau$ yields
\begin{equation}
    E_\ell(t_0 + \tau) = \sum_{i|m j} \epsilon_i n_{\ell i | m j} = \sum_i \epsilon_i n_{\ell i}(t_0 + \tau)
\end{equation}
as follows from \cref{eq:update}.

Assuming all microscopic realizations of each path in the system are equally probable, its macroscopic evolution $n_{IJ}$ maximizes the objective function
\begin{equation}\label{eq:obj}
    f(n_{IJ}) = \ln \Omega_t^{t_0+\tau} + \sum_\ell \beta_\ell B_\ell + \sum_J \theta_J A_J
\end{equation}
where $\theta_J$ and $\beta_\ell$ are the Lagrange multipliers to the constraint \cref{eq:cont_matter,eq:cont_energy} respectively. Applying the extremum conditions $\partial f / \partial n_{IJ} = 0$ and eliminating $\theta_J$, we then obtain
\begin{equation}\label{eq:transition}
    n_I(t_0 + \tau) = \sum_{J} P_{IJ}(\tau) n_J(t_0),
\end{equation}
in which we introduced a transition probability
$$
    P_{IJ} = \frac{w_{IJ}(\tau) \exp\big(-\beta_\ell(\epsilon_i - \epsilon_j)\big)}{\sum_{I'} w_{I'J}(\tau) \exp\big(-\beta_{\ell'} (\epsilon_{i'} - \epsilon_j)\big)},
$$
with weights $w_{IJ}(\tau) = g_I a_{IJ}(\tau) \gamma_{IJ}(\tau)$.

In addition, we assume that at a sufficiently small time scale $\tau\to0$ we can approximate
\begin{equation}\label{eq:Markov}
    \frac{d}{d\tau} P_{IJ}(\tau)\simeq K_{IJ},
\end{equation}
with constant rates $K_{IJ}$. Normalized by $N$, \cref{eq:transition} in the limit of $\tau\to0$ leads to a master equation for probabilities $p_I = n_I / N$ \SM{Sec.~\smMaster}:
\begin{equation}\label{eq:master}
    \frac{d}{dt} {p}_I= \sum_{J} K_{IJ} p_J.
\end{equation}
Numerical integration of \cref{eq:master} in time by an Euler step $\Delta t$ corresponds to the Markov chain
\begin{equation}\label{eq:chain}
    p_I(t + \Delta t) = \sum_J P_{IJ}(\Delta t) p_J(t)
\end{equation}
solved together with \cref{eq:cont_energy} for $\beta_\ell = (\kB T_\ell)^{-1}$, where $\kB$ is the Boltzmann constant and $T_\ell$ is a \textit{local} temperature defined for each lattice site.

To align a discrete energy structure $\epsilon_i$ with the collisional gas in the continuum limit in space $\Delta x \to 0$, $L \to \infty$, and time $\Delta t \to 0$, we parameterize the energy levels $\epsilon_i$ so that the particles' velocity at the level $i$ is given by
$$
    v_i = \lim_{\substack{\Delta x \to 0 \\ \Delta t \to 0}} i\, \Delta x / \Delta t,
$$ 
in which the index $i$ may be interpreted as the estimated maximum number of lattice sites a particle of given energy would be able to traverse in a time step $\Delta t$. Then the individual levels follow the structure $\epsilon_i = \varepsilon i^2 / 2$, where $\varepsilon$ is a constant parameter, aligned with the discretized kinetic energies $\mathcal{M} v_i^2 / 2$ of gas molecules of mass $\mathcal{M} \simeq \varepsilon (\Delta t / \Delta x)^2$.

The dynamical structure of \cref{eq:chain} can be specified by a graph, where nodes represent the energy levels $\epsilon_i$ distributed over the lattice sites $\ell$, and the edges are weighted by $w_{IJ}(\Delta t)$. In this letter we compare two such graphs.

First, we estimate effective, up to an arbitrary factor, weights $\hat{w}_{IJ}\propto w_{IJ}(\Delta t)$ and degeneracies $\hat{g}_I \propto g_I$ (\cref{fig:graph}) from the coarse-grained MD simulations of a hard-core gas~\SM{Sec.~\smDynamo}. We compute the coarse-grained states of the Boltzmann gas \textit{in equilibrium} from the atomistic simulation by constructing histograms of the particles' coordinates and energies, $(x_p, \mathcal{M} \bm{v}^2_p/2)_{p=1,2,... N}$. Then we measure the frequencies of transitions between consecutive frames of the simulation to find $\hat{g}_I$ and $\hat{w}_{IJ}$. This scheme, which we call \textit{kinetic connectivity} (KC), in principle admits asymmetric edge weights $\hat{w}_{IJ} \ne \hat{w}_{JI}$ over longer ranges $|\ell - m| > 1$.

In the second case, we consider the simplest nearest-neighbor (NN) graph connectivity \SM{Fig.~\sfigNN}, where all degeneracies are equal, $\hat{g}_I = g$, as traditionally used to analyze diffusive systems~\cite{Ghosh2006,companion}. In this case, the weights can be derived from the adjacency matrix
$$a_{\ell i | m j} = a_{\ell m} = a_{m\ell}= \begin{cases}
    \text{$1$ if $|\ell - m| \le 1$,}\\
    \text{$0$ otherwise,}
\end{cases}$$
limited to short-range transitions $|\ell - m| \le 1$. Assuming also that all paths are equivalent, we compute $\gamma_{IJ}  \in\{1,2\}$ by applying reflective boundary conditions, which avoid ``wall'' artifacts---density depletion at the end sites $\ell\in\{1,L\}$,---and thus ensure that the Boltzmann distribution is reproduced under equilibrium conditions~\SM{Sec.~\smMirror}.

Invoking the ensemble equivalence, we fix the Lagrange multipliers $\beta_1 = (k_B T_\lft)^{-1}$ and $\beta_L = (k_B T_\rgt)^{-1}$ of the terminal lattice sites $\ell\in\{1,L\}$. In the bulk of the system ($1 < \ell < L$) the continuity \cref{eq:cont_energy} determines $\beta_\ell$ as under the microcanonical conditions. The system thus exchanges heat with two reservoirs at temperatures $T_{\ell = 1,L} = (k_B \beta_\ell)^{-1}$ connected to the respective ends $\ell\in\{1,L\}$.

\paragraph*{Results.}---Under equilibrium conditions $T_\lft = T_\rgt$, numerical solutions~\cite{code} of \cref{eq:chain} with both the NN and KC dynamical schemes reproduce the bulk density and temperature profiles of the hard-core gas~\SM{Fig.~\sfigEq, Sec.~\smSim}. The latter scheme additionally captures the boundary effects of MD---not only at the temperature $T_\rf$ used in the calibration procedure.

\begin{figure}[t]
\includegraphics[width=\columnwidth]{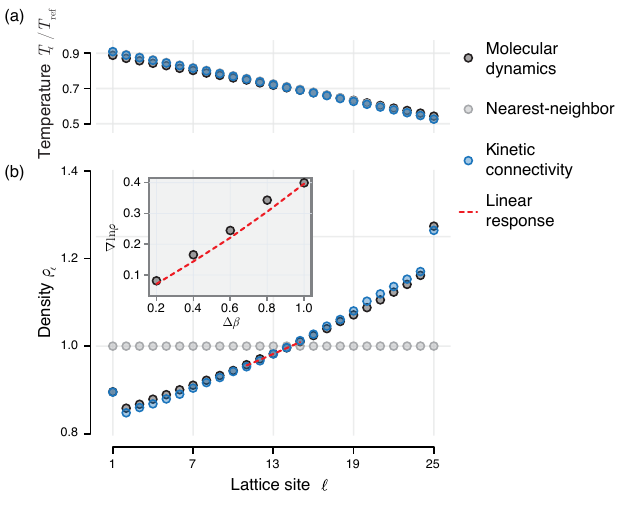}
\caption{\label{fig:ne}Profiles of temperature (a) and density (b) for a hard-core gas in a nonequilibrium steady state sustained by thermalizing walls at $\kB T_\lft=0.9$, $\kB T_\rgt=0.5$ (arb. u.). Our coarse-grained model with kinetic connectivity, calibrated from equilibrium data at $\kB T_\rf=1.0$, accurately predicts histograms of temperature and density in MD simulations. The nearest-neighbor scheme fails to reproduce the density distribution in the hard-core gas subject to the temperature drop. Linear-response formulas \SM{Eqs.~(\seqThermal)--(\seqSoret)} for the Soret coefficient $S_T$ and \cref{eq:trend} accurately describe the gradient $\nabla\ln\rho\vert_{\ell=13}\approx -S_T \nabla T$ over a wide range of differences $\Delta \beta = (\kB T_\rgt)^{-1} - (\kB T_\lft)^{-1}$ (inset).}
\end{figure}

Strikingly, no further calibration is required away from equilibrium. Although the KC model is parameterized only once under equilibrium conditions at $T_\rf$, it accurately predicts the steady-state temperature and density profiles of the hard-core gas under an imposed difference $\Delta T \ne 0$ (\cref{fig:ne}). The agreement remains quantitative across the entire system including boundary effects, even far from equilibrium $T_\rgt \ll T_\lft \lesssim T_\rf$ \SM{Table \stabArgon}. By contrast, the NN dynamics generates flat density profiles despite the applied temperature gradient, emphasizing the role of the transition-graph structure for coupling thermal and diffusive transport.

The network structure we obtain for the KC dynamics reveals two main contributions to the phenomenon of thermodiffusion. First, the multiplicities $\hat{\gamma}_{IJ}$ of long-range transitions $(\ell i | m j): |\ell - m| > 1$ increase with the source energy level $j$, i.e. higher source energies allow particles to traverse large spatial distances via a greater number of distinct paths compared to particles with lower source energies [\cref{fig:graph}(c)]. Second, such large jumps are typically accompanied by a loss of kinetic energy, i.e. a transition to a lower level $i < j$~[\cref{fig:graph}(d)]. Combined with the condition of local equilibrium, which populates high-energy states at the hot end of the system, these two aspects favor faster migration of particles to the cold side where they accumulate in steady state.

As the Boltzmann gas quickly converges to the conditions of local equilibrium, the distributions $p_{i|\ell}(t)$ over energy levels $i=1,2,...,M$ at each site $\ell$ become canonical $\bar{p}_{i|\ell}(\beta_\ell) \propto g_{\ell i} \exp(-\beta_\ell \epsilon_i)$. This property is not an assumption, but a fact numerically verified by calculating the Kullback-Leibler divergence between $p_{i|\ell}(t)$ and $\bar{p}_i(\beta_\ell)$~\SM{Fig.~\sfigKL}.

From the condition of local equilibrium, by leveraging the linear-response theory, we can derive the thermodiffusion equation~\SM{Sec. \smAnalysis}, including closed-form Eqs.~(\seqThermal)--(\seqSoret) for $D$ and $D_T$ in \cref{eq:no_diffusion}, and the Soret coefficient $S_T$ \cite{Parola2004,Burelbach2018,Duhr2006PRL,Duhr2006PNAS,Brenner2010}. The density profile in the bulk then follows the trend predicted from~\cref{eq:no_diffusion} [\cref{fig:ne}(b)]:
\begin{equation}\label{eq:trend}
\rho \propto \exp(-S_T \nabla T x).
\end{equation}

\paragraph*{Discussion.}---Our graph representation of microscopic dynamics quantitatively predicts nonequilibrium steady states from equilibrium statistics of the hard-core gas. This approach exposes the key mechanism of thermodiffusion that consists in coupling thermal and diffusive transport through the available transitions between molecular states.

When the local distribution of particles over energy levels becomes inhomogeneous due to the temperature gradient, heat and matter are exchanged through distinct, inequivalent paths in the forward and reverse directions of the flows. In the forward direction the particles have more options of long-range transitions and dissipate their energy, whereas in the reverse they choose among realizations of shorter moves and carry less energy. Thereby the transport coefficients of thermodiffusion acquire a statistical interpretation in terms of the multiplicities of microscopic transition paths.

The quantitative predictive agreement of the KC model with molecular dynamics across the entire system, including boundary layers and far-from-equilibrium conditions, shows that the graph structure contains the essential microscopic information governing thermodiffusion. By contrast, nearest-neighbor dynamics are insufficient to capture microscopic effects coupling energy and matter currents. In this scheme the number of viable long-range transitions neither increases with a particle's energy nor is biased toward lower target energies.

More generally, representing microscopic dynamics as a graph of allowed transitions provides a statistical description of transport whose calibration links equilibrium fluctuations to nonequilibrium steady states. We expect this framework to extend to other transport phenomena governed by constrained microscopic dynamics.

\begin{acknowledgments}
The authors acknowledge funding from the EMBL. JE was supported in part by HFSP grant (reference no. RGP007/2023). The authors acknowledge the use of artificial-intelligence tools~\cite{chatgpt,codex} for bibliographic search, as well as for language and coding assistance. The generated code was reviewed, manually refined, commented, and documented. The authors take full responsibility for the scientific content, interpretation, and conclusions of the manuscript. The authors are also grateful to Ian Estabrook, Patrick Jentsch, Samuele Massimi, Michele Lupini, and Charalampos Pozoukidis for their feedback.
\end{acknowledgments}

\section*{Data Availability Statement}
The data that support the findings of this study were generated by numerical simulations. The source code and parameters used to generate the simulations is publicly available~\cite{SM,code}.

\bibliography{main}

\end{document}